\documentclass[a4paper,11pt]{article}
\usepackage{pos}
\usepackage{caption}
\usepackage{subcaption}
\renewcommand{\logo}{\relax}

\title{Jet-Flavour Tagging at FCC-ee}

\author*[a, b, 1]{Kunal Gautam}

\affiliation[a]{Inter-university Institute for High Energies, Vrije Universiteit Brussel,\\
1050 Brussels, Belgium}

\affiliation[b]{Universit\"at Z\"urich,\\
8057 Zurich, Switzerland}

\note{On behalf of the FCC collaboration}

\emailAdd{kunal.gautam@cern.ch}

\abstract{Jet-flavour identification algorithms are of paramount importance to maximise the physics potential of the Future Circular Collider (FCC).
Out of the extensive FCC-ee physics program, flavour tagging is crucial for the Higgs physics program, given the dominance of hadronic decays of the Higgs boson. Highly efficient discrimination of $b$-, $c$-, $s$-, and gluon jets allows access to novel decay modes that cannot be identified at the LHC, adding quantitatively new dimensions to the Higgs physics programme.

This contribution presents new jet flavour identification algorithms based on advanced machine-learning techniques that exploit particle-level information. Beyond an excellent performance of $b$- and $c$-quark tagging, they are also able to discriminate jets from strange quark hadronisation, opening the way to improve the sensitivity of the Higgs to strange quark coupling. The impact of different detector design assumptions on the flavour tagging performance is assessed using one of the baseline detector concepts for FCC-ee, IDEA.}

\FullConference{%
  41st International Conference on High Energy physics - ICHEP2022\\
  6-13 July, 2022\\
  Bologna, Italy
}

\begin{document}
\maketitle

\section{Introduction} \label{sec:intro}
One of the main objectives of FCC-ee \cite{Benedikt:2651299} is the precise measurements of Standard Model parameters, like the couplings of the Higgs boson to the bottom and charm quarks and gluons. This requires an efficient reconstruction and identification of the hadronic final states of these processes, which entails identifying the flavour of the parton that initiated the jet, referred to as jet-flavour tagging. Efficient and accurate jet-flavour identification is also necessary to assess the feasibility of measurements such as $Z \rightarrow s\bar{s}$ or $H \rightarrow s\bar{s}$ and therefore is essential to utilise the maximal physics potential of future collider experiments. Another aim of the jet-flavour tagging studies is to drive the detector development by providing requirements that result in improved performance in identifying the hadronic final states, which will reduce the statistical limits on physics measurements and will open up the potential to access previously unobserved or imprecisely-measured physics channels.

Hadronic jets originating from the heavier $b$ and $c$ quarks contain $b$ and $c$ hadrons. These have a significant lifetime and tend to decay at some distance from the interaction point. The charged tracks can be clustered to reconstruct these displaced decay vertices, known as secondary vertices (SVs). These SVs can be used to distinguish the heavier quark jets from the lighter quark jets, which typically don't have any SVs. The strange jets tend to have a higher (lower) multiplicity of Kaons (Pions) than up/down jets, which consist the majority of the background while tagging s-jets. Therefore, particle identification (PID) techniques that can distinguish $K^{\pm}$ and $\pi^{\pm}$ and can reconstruct $K^{0}_{S}$ are crucial for $s$-jet tagging.

This contribution reports on three different algorithms that use advanced machine-learning techniques to identify hadronic jet flavours. Two of the algorithms have been developed for FCC-ee and are motivated by tagging algorithms currently being used at the LHC experiments and one algorithm has been developed for the International Linear Collider (ILC) \cite{Adolphsen:1601969}.
The effects of different detector designs and of using different input variables are also discussed.

\section{Flavour Tagging with Transformer-based Neural Network Architecture} \label{sec:transformer}
This tagging algorithm using a novel transformer-based neural network architecture that relies on the attention mechanism \cite{https://doi.org/10.48550/arxiv.1706.03762} has been adopted for use in the FCC-ee case. The network was trained using samples of $e^{-}e^{+} \rightarrow Z \rightarrow q\bar{q}$ process, where $q \equiv u, d, s, c, b$, at the center-of-mass energy ($\sqrt{s}$) of $91.2$ GeV. 
\texttt{Pythia8.303} \cite{https://doi.org/10.48550/arxiv.2203.11601} is used for the parton showering and hadronisation, event reconstruction is done with \texttt{Delphes} \cite{de_Favereau_2014} using fast-simulation of the IDEA detector concept \cite{Bedeschi:2021bS}, and jet clustering is performed with $e^{-}e^{+}$ $k_{T}$ algorithm \cite{Catani:1991hj} using \texttt{FastJet-3.3.4} \cite{Cacciari_2012}.

The stable jet constituents are distributed in particle-flow categories, charged (neutral) hadrons, electrons, muons, and photons. The charged tracks are used to reconstruct V$^{0}$ vertices to identify $K^{0}_{S}$ and $\Lambda^{0}$, which helps to improve the $s$-tagging performance. The remaining tracks are clustered to reconstruct SVs, which are relevant for $b$- and $c$-tagging. Vertex reconstruction is performed using an implementation of the vertexing module of the \texttt{LCFIPlus} framework \cite{Suehara_2016}.

\begin{figure}
    \centering
    \begin{subfigure}[b]{0.48\textwidth}
        \centering
        \includegraphics[width=\textwidth]{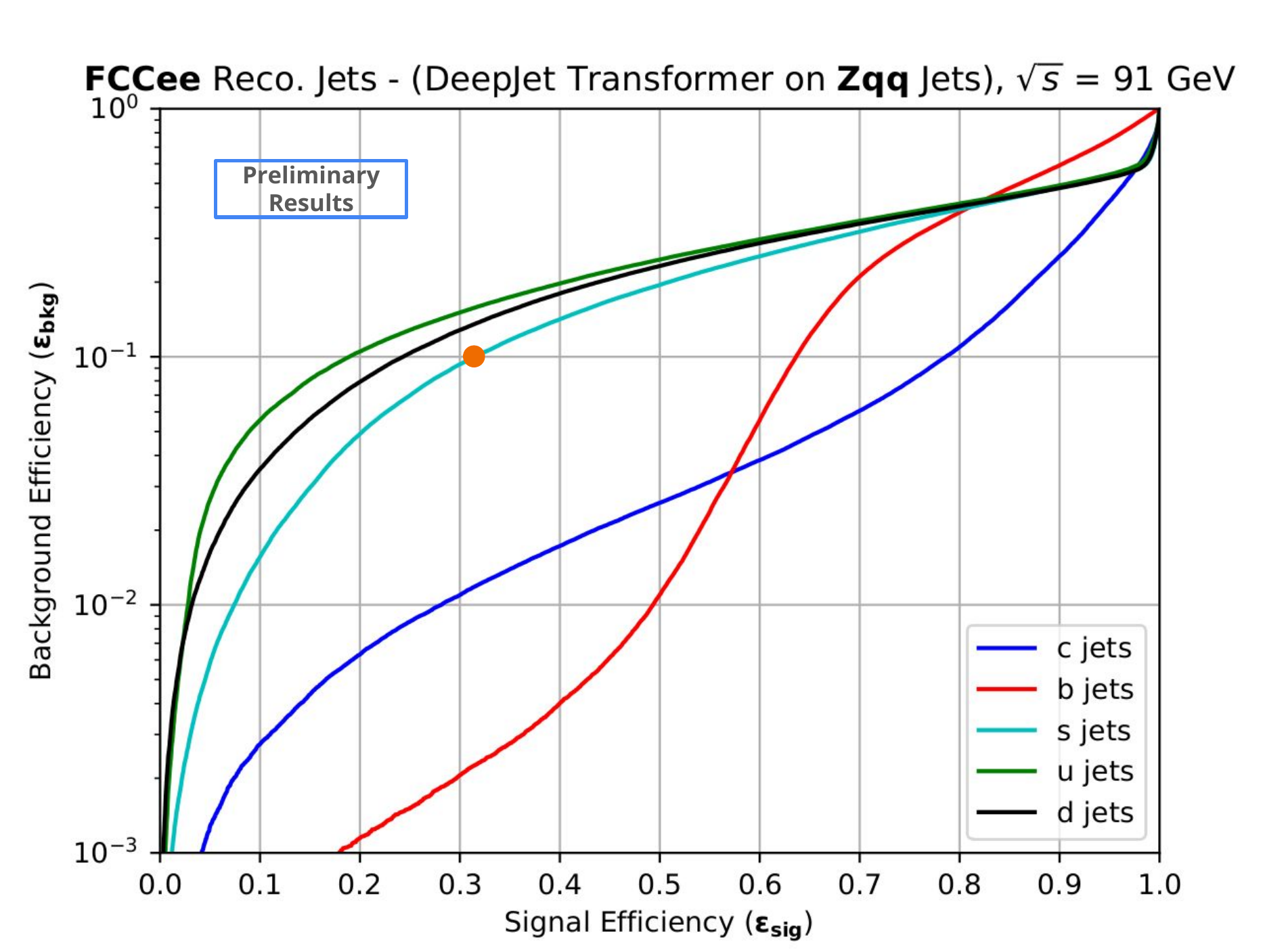}
        \caption{Performance without V$^0$s}
        \label{fig:v0_no}
    \end{subfigure}
    \hfill
    \begin{subfigure}[b]{0.48\textwidth}
        \centering
        \includegraphics[width=\textwidth]{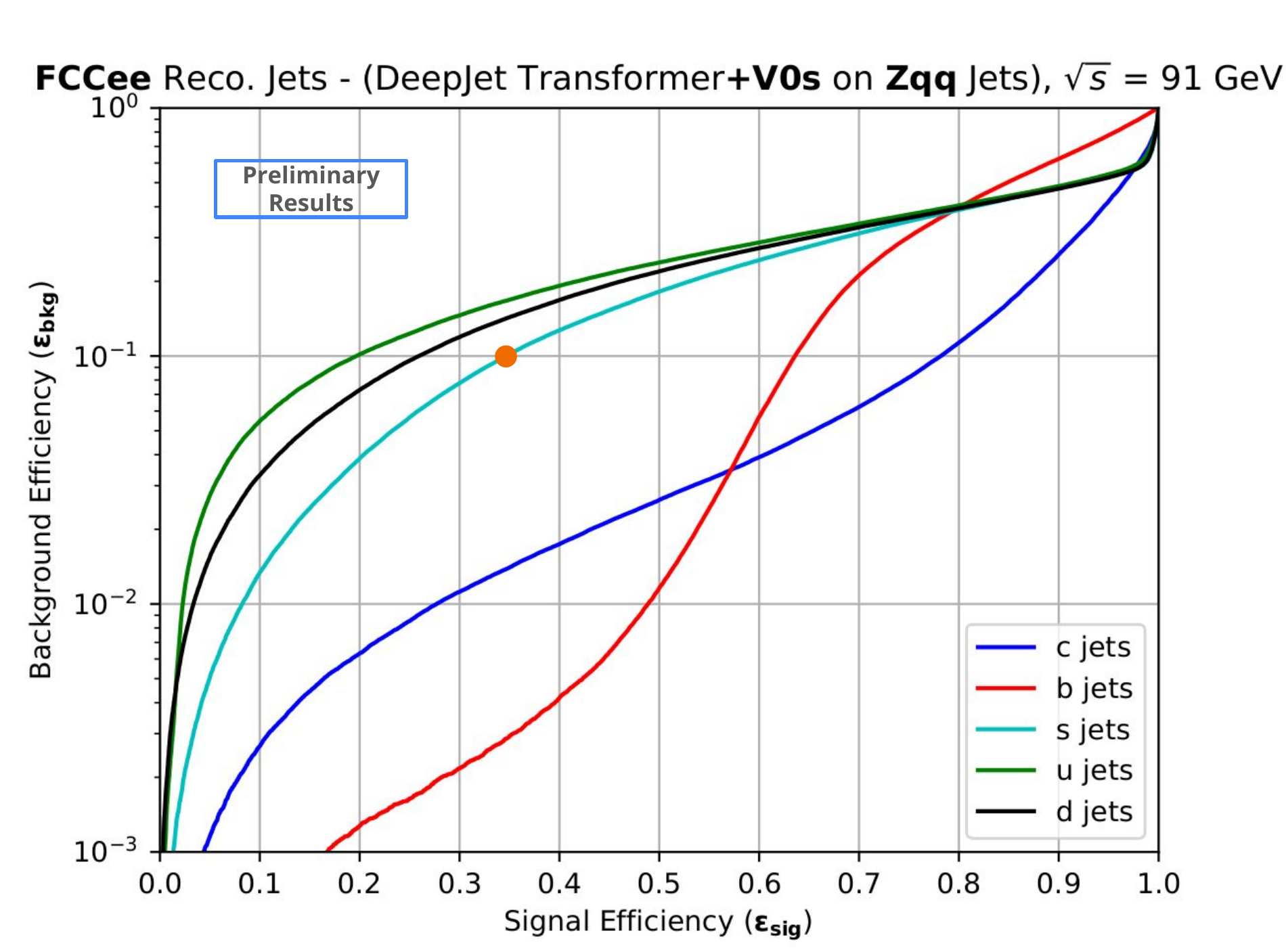}
        \caption{Performance with V$^0$s}
        \label{fig:v0_yes}
    \end{subfigure}
    \caption{Impact of V$^0$s on performance of jet-flavour tagging with transformer-based neural network.}
    \label{fig:transformer1}
\end{figure}

The orange point in figure \ref{fig:transformer1} shows that the performance of $s$-tagging improves by a few percent by adding V$^0$s. While identifying $s$-jets, the background consists mainly of $u$- and $d$-jets and, to reduce it, PID variables are crucial.

\begin{figure}[b]
    \centering
    \begin{subfigure}[b]{0.48\textwidth}
        \centering
        \includegraphics[width=\textwidth]{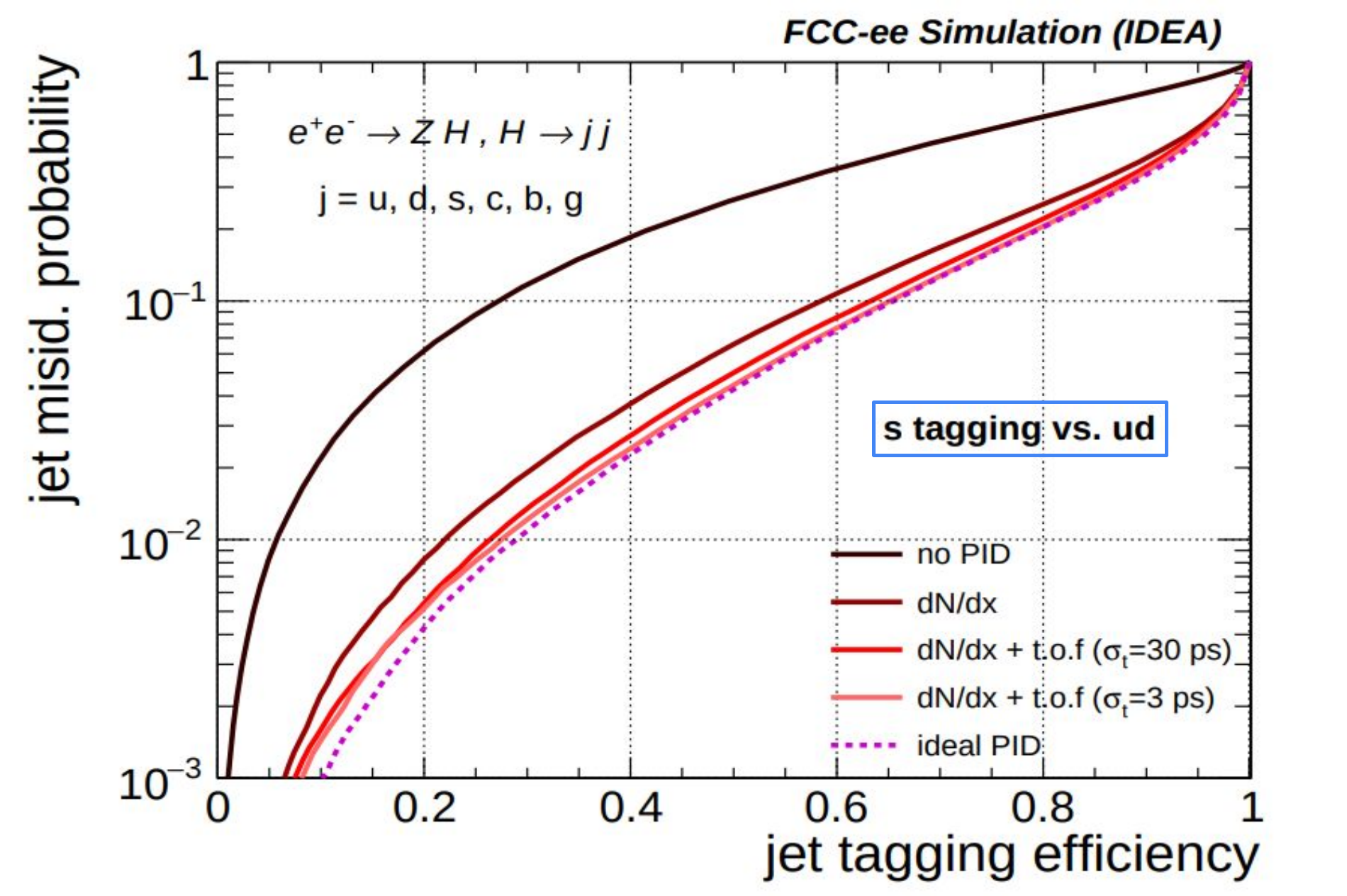}
        \caption{Impact of PID on $s$-tagging}
        \label{fig:gnn_s}
    \end{subfigure}
    \hfill
    \begin{subfigure}[b]{0.48\textwidth}
        \centering
        \includegraphics[width=\textwidth]{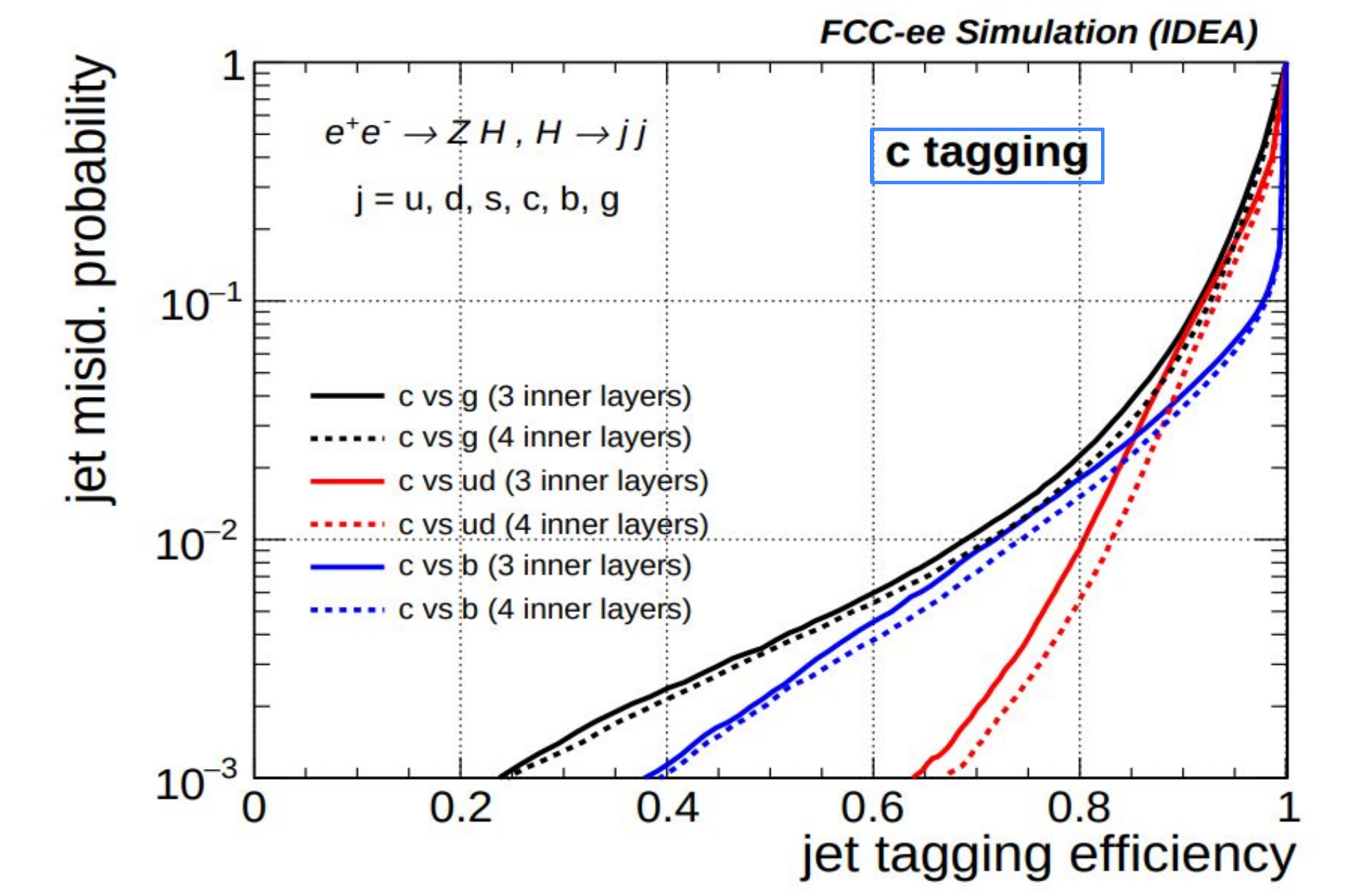}
        \caption{Impact of inner tracking geometry on $c$-tagging}
        \label{fig:gnn_c}
    \end{subfigure}
    \caption{Effect of detector designs on performance of jet-flavour tagging with GNN \cite{Bedeschi_2022}.}
    \label{fig:gnn1}
\end{figure}

\section{Flavour Tagging with Graph Neural Network Architecture} \label{sec:gnn}
\texttt{ParticleNetIDEA} \cite{Bedeschi_2022} is a tagging algorithm based on graph neural networks (GNN) and is adopted from \texttt{ParticleNet} jet-tagging algorithm \cite{Qu_2020}. 
The simulated sample used to train the network consists of $e^{-}e^{+} \rightarrow ZH$ events produced at the $\sqrt{s}=240$ GeV with $H \rightarrow gg/q\bar{q}$, where $q \equiv u, d, s, c, b$. The $Z$ boson is forced to decay to a neutrino pair. The training is performed with five different samples, corresponding to each jet-flavour category ($ud, s, c, b, g$), containing $10^{6}$ events each. \texttt{Pythia8} \cite{https://doi.org/10.48550/arxiv.2203.11601} is used for the decay, parton shower, and hadronisation; final state particles are reconstructed using \texttt{Delphes PF} algorithm. PID variables, number of ionisation clusters ($dN/dx$) and time-of-flight, are calculated using dedicated modules in \texttt{Delphes} \cite{de_Favereau_2014}. The IDEA detector concept \cite{Bedeschi:2021bS} is used. Jet clustering is performed with generalised $e^{-}e^{+}$ $k_{T}$ algorithm \cite{Catani:1991hj} with parameter $p=-1$, using \texttt{FastJet-3.3.4} \cite{Cacciari_2012}.

Three sets of input variables are used for training: kinematic variables derived from the momentum of each jet-constituent, displacement variables related to the longitudinal and transverse displacement of charged tracks (relevant for $b$- and $c$-tagging), and identification variables derived from PF reconstruction and PID algorithms.

At $90\%$ efficiency in $b$-tagging, there is $2\%$ misidentification rate for $g$- and $c$-jets. At $90\%$ efficiency in $c$-tagging, there is a $7\%$ misidentification rate for $g$ and $ud$-jets and $4\%$ for $b$-jets.
The impact of PID on $s$-tagging performance can be seen in Figure \ref{fig:gnn_s}. Cluster counting brings in the most gain and, along with time-of-flight measurement with $30$ ps timing resolution, performs very close to the scenario with perfect PID. Figure \ref{fig:gnn_c} shows that $2$x background rejection, consisting of $ud$-jets, can be achieved in $c$-tagging with an additional pixel layer in the inner tracker.

\section{Flavour Tagging with Recurrent Neural Network Architecture} \label{sec:rnn}
\begin{figure}
    \centering
    \includegraphics[scale=0.4]{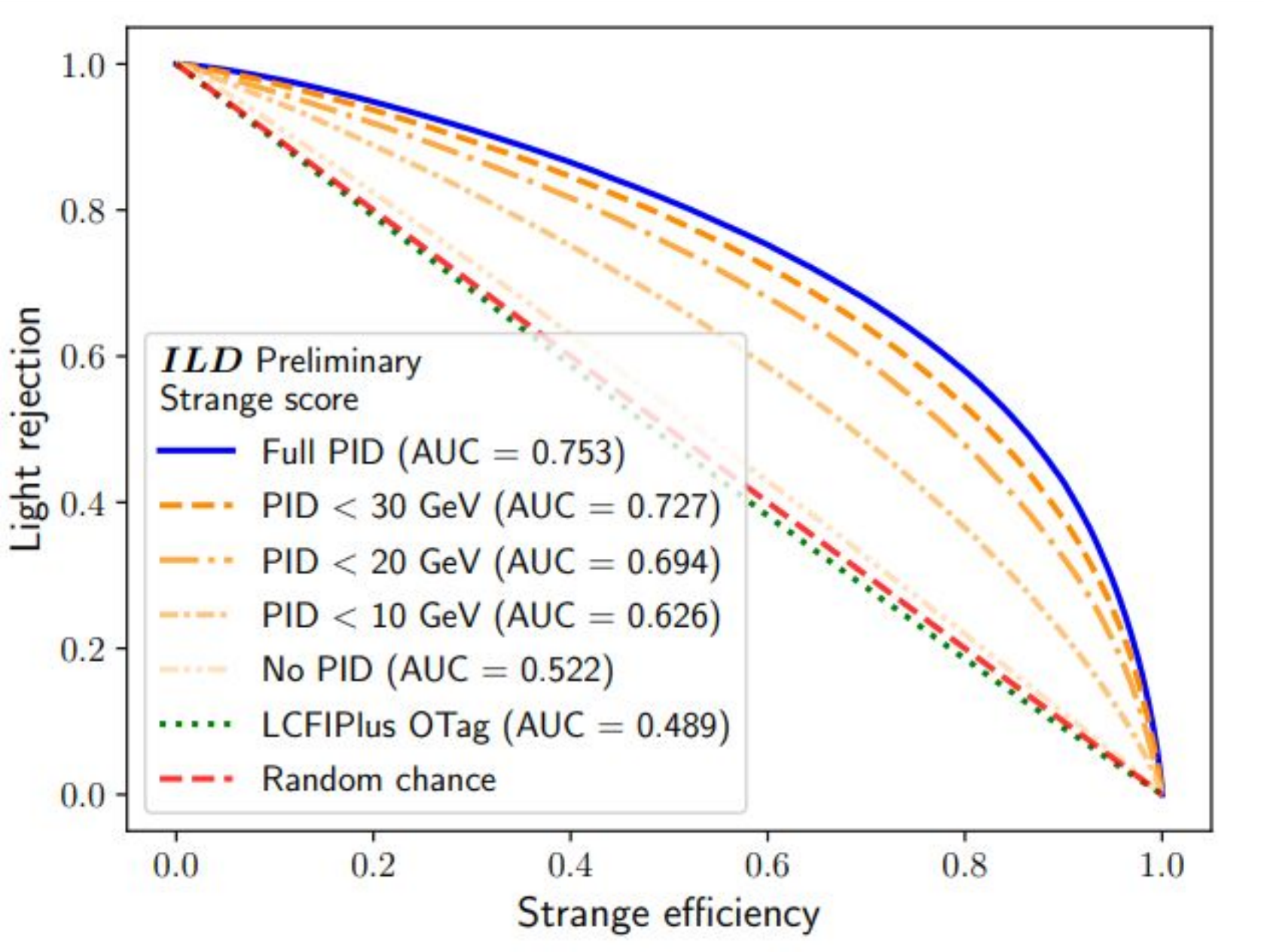}
    \caption{Comparison of $s$-tagging performance of jet-flavour tagging with RNN for PID in different momentum ranges \cite{https://doi.org/10.48550/arxiv.2203.07535}.}
    \label{fig:rnn1}
\end{figure}

A tagging algorithm \cite{https://doi.org/10.48550/arxiv.2203.07535} using recurrent neural networks (RNN) developed for the ILC also substantiates the results of the taggers developed for FCC-ee. The algorithm reports significant improvement in $s$-tagging performance compared to the ILC baseline tagger \cite{Suehara_2016} that does not use any PID information. Figure \ref{fig:rnn1} shows the effect of PID information for different momentum range on the $s$-tagging performance. Due to such crucial importance of PID, a compact RICH detector has also been proposed for ILC \cite{https://doi.org/10.48550/arxiv.2203.07535}.

\section{Conclusion} \label{sec:conclusion}
The state-of-the-art jet-flavour tagging algorithms being developed for FCC-ee show promising results and have excellent performance in $b$- and $c$-tagging modes. Additional inner-tracking layers or reaching closer to the interaction point will improve the performance further. It has motivated studying the feasibility of a smaller beam pipe.
The clean environment at FCC-ee and the precisely known initial state, along with the use of advanced ML algorithms, make possible new analysis techniques like $s$-tagging. A good PID strategy is necessary for a well-performing $s$-tagger. Cluster counting, time-of-flight, and V$^0$ reconstruction provide good PID for particles in the momentum range of interest. Suitably designed detectors and algorithms that can exploit the full potential of these detectors will enable future colliders like FCC-ee to study rare processes and maximally utilise their extensive dataset to reach their goal of high-precision measurements.

\acknowledgments{I want to thank and acknowledge contributions from M. Basso, F. Bedeschi, F. Blekman, V. Cairo, F. Canelli, A. De Moor, A. Macchiolo, L. Gouskos, A. Ilg, E. Plörer, and M. Selvaggi.

The author is supported by FWO (Belgium) and SNF (Switzerland).}

\bibliographystyle{JHEP}
\bibliography{skeleton}

%

\end{document}